\documentclass[nonatbib]{ragtime} 

\title[Electro-vacuum fields around black holes]%
      {
      Stationary electro-vacuum fields around black holes}

\author[V.~Karas]  
       {Vladim\'{\i}r Karas\\ 
        \ins{1}Astronomical~Institute, Academy~of~Sciences, Bo\v{c}n\'{\i}~II~1401,\splitins[1]
        CZ-14100~Prague, Czech~Republic\\
        \ins{}E-mail:~\Email{vladimir.karas@cuni.cz} 
}
\begin{document}

\newcommand{\beq}{\begin{equation}}
\newcommand{\eeq}{\end{equation}}
\newcommand{\rd}{{\,\rm d}}
\newcommand{\ri}{{\rm{i}}}
\newcommand{\nab}{\mbox{\protect\boldmath$\nabla$}}
\renewcommand{\vec}[1]{\mbox{\protect\boldmath$#1$}}

\begin{abstract}
This is the second lecture of `RAGtime' series on electrodynamical
effects near black holes. We will summarize the basic equations of
relativistic electrodynamics in terms of spin-coefficient
(Newman-Penrose) formalism. 

The aim of the lecture is to present
important relations that hold for exact electro-vacuum solutions and to
exhibit, in a pedagogical manner, some illustrative solutions and useful
approximation approaches. First, we concentrate on weak electromagnetic fields
and we illustrate their structure by constructing the magnetic and electric
lines of force. Gravitational field of the black hole assumes axial symmetry, whereas 
the electromagnetic field may or may not share the same symmetry.
With these solutions we can investigate the frame-dragging
effects acting on electromagnetic fields near a rotating black hole.
These fields develop magnetic null points and current sheets. Their
structure suggests that magnetic reconnection takes place near the
rotating black hole horizon. Finally, the last section will be devoted
to the transition from test-field solution to exact solutions of coupled
Einstein-Maxwell equations. 

New effects emerge within the framework 
of exact solutions: the expulsion of the magnetic flux out of the black-hole
horizon depends on the intensity of the imposed magnetic field.
\end{abstract}


\section{Introduction}
Electromagnetic fields play an important role in astrophysics. Near
rotating compact bodies, such as neutron stars and black holes, the
field lines are deformed by an interplay of rapidly moving plasma and
strong gravitational fields. Here we will illustrate purely
gravitational effects by exploring simplified vacuum solutions in which
the influence of plasma is ignored but the presence of strong gravity is
taken into account.

In the first lecture of this workshop series (Karas 2005, Paper~I) we 
summarized the basic equations of relativistic magnetohydrodynamics
(MHD). In that paper we employed standard tensorial notation and we
focused our attention on situations when the plasma motion is governed
by MHD and gravitational effects are competing with each other in the
vicinity of a black hole. We limited our discussion to axially symmetric
and stationary flows. The latter assumption will be still maintained in
the present talk. In fact, we will restrict ourselves to purely
electro-vacuum solution, however, we will discuss them in greater depth
and, more importantly, we will employ the elegant formalism of null
tetrads.  We do not derive new solutions or technique in these lectures,
instead, we summarise useful relations in the form of brief notes
paying special attention to effects of strong gravity. 

One new point is mentioned in conclusion: with {\em exact solutions\/} 
of Einstein-Maxwell electrovacuum fields, an aligned magnetic flux becomes 
expelled from a rotating black hole as an interplay between the shape of
magnetic lines of force (which become pushed out of the horizon) and
the concentration of the magnetic flux tube toward the rotation axis (which 
becomes more concentrated for strong magnetic fields because of their
own gravitational effect). This is, however, important only for {\em very strong\/}
magnetic fields only, where `very strong' means that the magnetic field contributes
to the space-time metric.

\section{Definitions, notation, and basic relations}
\subsubsection*{Field equations}
We start with Einstein's equations which, in the notation
of Paper~I, take a familiar form of a set of coupled 
partial differential equations (e.g.\ Chandrasekhar 1983),
\beq 
R_{\mu\nu}-\textstyle{\frac{1}{2}}Rg_{\mu\nu}=8\pi T_{\mu\nu},
\eeq
where the right-hand side source terms $T_{\mu\nu}$ are of purely
electromagnetic origin,
\beq
T^{\alpha\beta}\equiv T^{\alpha\beta}_{\rm EMG}=\frac{1}{4\pi}\left(F^{\alpha\mu}F^\beta_\mu- \frac{1}{4}F^{\mu\nu}F_{\mu\nu}g^{\alpha\beta}\right),
\eeq
\beq
{T^{\mu\nu}}_{;\nu}=-F^{\mu\alpha}j_{\alpha},\qquad
{F^{\mu\nu}}_{;\nu}=4\pi j^\mu,\qquad {^\star F^{\mu\nu}}_{;\nu}=4\pi\mathcal{M}^\mu.
\eeq
where $^\star
F_{\mu\nu}\equiv\frac{1}{2}{\varepsilon_{\mu\nu}}^{\rho\sigma}F_{\rho\sigma}$.
We assume that the electromagnetic test-fields reside in a curved
background of a  rotating black hole. Such solutions can be found by
solving for the electromagnetic  field in a fixed background geometry of
Kerr metric (Thorne et al.\ 1986; Gal'tsov 1986). Here we study classical
solutions for (magnetised) Kerr-Newman black holes that possess a horizon. 
Higher-dimensional black holes and black rings in external magnetic fields were explored
by, e.g., Ortaggio (2005), Yazadjiev (2006), and references cited therein, whereas
an extension to the case of naked singularity has been discussed recently by 
Ad\'amek \& Stuchl\'{\i}k (2013).

\subsubsection*{Killing vectors generate a test-field solution}
The presence of Killing vectors corresponds to the symmetry of the
spacetime (Chandrasekhar 1983; Wald 1984), such as stationarity and
axial symmetry.  Killing vectors satisfy the well-known equation,
\beq
\xi_{\mu;\nu}+\xi_{\nu;\mu}=0,
\eeq
where coordinate system is selected in such a way that the following
condition is satisfied: 
$\xi^\mu=\delta^\mu_\rho$. 
One can check that Killing vectors obey a sequence of relations:
\beq
0=\xi_{\mu;\nu}+\xi_{\nu;\mu}=\xi_{\mu,\nu}-\Gamma^\lambda_{\mu\nu}\xi_\lambda+\xi_{\nu,\mu}-\Gamma^\lambda_{\mu\nu}\xi_{\lambda}=g_{\mu\nu,\rho}.
\label{kill}
\eeq
The last equality (\ref{kill}) states that because of symmetry the metric 
tensor does not depend $x^\rho$ coordinate.

The electromagnetic field may or may not conform to the same symmetries
as the gravitational field. Naturally, the problem is greatly simplified
by assuming axial symmetry and stationarity for both fields.
In a vacuum spacetime, Killing vectors generate a  test-field solution
of Maxwell equations. We \textit{define} the electromagnetic field by
associating it with the Killing vector field,
\beq
F_{\mu\nu}=2\xi_{\mu;\nu}.
\eeq
Then
\beq
F_{\mu\nu}=2\xi_{\mu;\nu}=-2\xi_{\nu;\mu}=-F_{\nu\mu},
\eeq
\beq
F_{\mu\nu}=\xi_{\mu;\nu}-\xi_{\nu;\mu}\equiv\xi_{[\mu;\nu]}.
\eeq

By employing the Killing equation and the definition of Riemann tensor, i.e., the relations
$\xi_{\mu;\nu;\sigma}-\xi_{\mu;\sigma;\nu}=-R_{\lambda\mu\nu\sigma}\xi^\lambda$, and
$R_{\lambda[\mu\nu\sigma]\mathrm{cycl}}=0$,
we find:
\beq
\xi_{\mu;\nu;\sigma}=R_{\lambda\sigma\mu\nu}\xi^\lambda,\qquad
{\xi^{\mu;\nu}}_{;\nu}={R^\mu}_\lambda\xi^\lambda.
\eeq
The right-hand side vanishes in vacuum, hence
\beq
{F^{\mu\nu}}_{;\nu}=0.
\eeq
It follows that the well-known field invariants are given by relations
\beq
\mbox{\boldmath$E. B$}=\textstyle{\frac{1}{4}}\,{^\star}F_{\mu\nu}F^{\mu\nu},
\qquad
B^2-E^2=\textstyle{\frac{1}{2}}F_{\mu\nu}F^{\mu\nu}.
\eeq

\subsubsection*{Magnetic and electric charges}
We start from the axial and temporal Killing vectors, existence of which
is guaranteed in any axially symmetric and stationary spacetime,
\beq
\xi^\mu=\frac{\partial}{\partial t},\qquad \tilde{\xi}^\mu=\frac{\partial}{\partial \phi}.
\eeq
In the language of differential forms (e.g.\ Wald 1984),
\beq
\underbrace{\textstyle{\frac{1}{2}}F_{\mu\nu}\rd x^\mu\wedge\rd x^\nu}_{\mbox{${\bf F}$}}=\underbrace{\xi_{\mu,\nu}\rd x^\mu\wedge\rd x^\nu}_{\mbox{${\bf d}${\protect\boldmath$\xi$}}}.
\eeq
The above-given equations allow us to introduce the magnetic and electric charges in the
form of integral relations,
\begin{eqnarray}
\mbox{Magnetic~charge:~}\qquad4\pi \mathcal{M}=\int_\mathcal{S}{\bf F}&=&\int_\mathcal{S}\mbox{${\bf d}${\protect\boldmath$\xi$}} =0.\\
\mbox{Electric~charge:}\qquad4\pi Q=\int_\mathcal{S}\mbox{\protect\boldmath$^\star F$}&=&\int_\mathcal{S}\mbox{{\protect\boldmath$^\star$}${\bf d}${\protect\boldmath$\xi$}}=-8\pi M,\\
&=&\int_\mathcal{S}\mbox{{\protect\boldmath$^\star$}${\bf d}${\protect\boldmath$\tilde{\xi}$}}=16\pi J,
\end{eqnarray}
where $M$ has a meaning of mass and $J$ is angular momentum of the source.
Here, integration is supposed to be carried out far from the 
source, i.e.\ in spatial infinity of Kerr metric in our case. 
For example for the electric charge we obtain
\beq
4\pi Q=\int_\mathcal{S}\mbox{\protect\boldmath$^\star F$}=\int_\mathcal{S}\,^{\star}F_{\mu\nu}\rd\sigma^{\mu\nu}=\int_{\mathcal{V}}2{F^{\tau\alpha}}_{;\alpha}\rd\mathcal{V},
\eeq
where
$\rd\sigma^{\mu\nu}=\rd_1x^\mu\wedge\rd_2x^\nu=\rd\theta\,\rd\phi$.

\subsubsection*{Wald's field}
In an asymptotically flat spacetime, $\partial_\phi$ generates uniform
magnetic field, whereas the field vanishes asymptotically for
$\partial_t$. These two solutions are known as the Wald's field (Wald
1974; King et al.\ 1975; Bi\v{c}\'ak \& Dvo\v{r}\'ak 1980;  
Nathanail \& Contopoulos 2014):
\beq
F=\textstyle{\frac{1}{2}}B_0\left(\rd\tilde{\xi}+\frac{2J}{M}\rd\xi\right).
\eeq
Magnetic flux surfaces:
\beq
4\pi \Phi_{\mathcal{M}}=\int_\mathcal{S}\mbox{\protect\boldmath$F$}\;=\;\mbox{const}.
\eeq
Magnetic and electric Lorentz force are then given by equations
\beq
m\mbox{\boldmath$\dot{u}$}=q_{\rm{m}}\mbox{\boldmath${^\star}F.u$},\qquad m\mbox{\boldmath$\dot{u}$}=q_{\rm{e}}\mbox{\boldmath$F.u$}.
\eeq
Finally, magnetic field lines (in the axisymmetric case):
\beq
\frac{{\rd{r}}}{{\rd\theta}}=\frac{B_r}{B_\theta},
\label{l1}
\eeq
Magnetic field lines lie in surfaces of constant magnetic flux (see below).

\section{Spin-coefficient formalism of null tetrads for electromagnetic fields}
The spin-coefficient formalism (Newman \& Penrose 1962) is a special case of 
the tetrad formalism where
tensors are projected onto a complete vector basis at each point in spacetime,
The vector basis is chosen as a complex null tetrad, 
$l^\mu$, $n^\mu$, $m^\mu$, $\bar{m}^\mu$, satisfying conditions
\beq
l_\nu n^\nu=1,\qquad m_\nu\bar{m}^\nu=-1,
\eeq
and zero all other combinations. A natural correspondence with an orthonormal tetrad reads
\beq
e_{(0)}=\frac{l+n}{\sqrt{2}},\quad
e_{(1)}=\frac{l-n}{\sqrt{2}},\quad
e_{(2)}=\frac{m+\bar{m}}{\sqrt{2}},\quad
e_{(3)}=\frac{m-\bar{m}}{\Im\,\sqrt{2}}.
\eeq
Null tetrads are not unambiguous, as the following three transformations maintain
the tetrad properties:
\begin{description}
\item[(i)]~$l\rightarrow l$, $m\rightarrow m+al$, $n\rightarrow n+a\bar{m}+\bar{a}m+a\bar{a}l$;\\[-3pt]
\item[(ii)]~$n\rightarrow n$, $m\rightarrow m+bm$, $l\rightarrow l+b\bar{m}+\bar{b}m+b\bar{b}n$;\\[-3pt]
\item[(iii)]~$l\rightarrow \zeta l$, $n\rightarrow \zeta^{-1}l$, $m\rightarrow e^{\Im\psi}m$;
\end{description}
with $\zeta$, $\psi\in\Re$.

Instead of six real components of $F_{\mu\nu}$, the framework of the null tetrad formalism describes the electromagnetic field by three independent complex quantities,
\begin{eqnarray}
\Phi_0&=&F_{\mu\nu}l^\mu m^\nu,\\
\Phi_1&=&\textstyle{\frac{1}{2}}F_{\mu\nu}\left(l^\mu n^\nu+\bar{m}^\mu m^\nu\right),\\
\Phi_2&=&F_{\mu\nu}\bar{m}^\mu n^\nu.
\end{eqnarray}
It can be checked that the backward transformation has a form
\beq
F_{\mu\nu}=\Phi_1\left(n_{[\mu}l_{\nu]}+m_{[\mu}\bar{m}_{\nu]}\right)+\Phi_2l_{[\mu}m_{\nu]}+\Phi_0\bar{m}_{[\mu}n_{\nu]}+c.c.
\eeq
The Newman-Penrose formalism defines the following differential 
operators:
\beq
D\equiv l^\mu\partial_\mu,\quad \delta\equiv m^\mu\partial_\mu,\quad \bar{\delta}\equiv\bar{m}^\mu\partial_\mu,\quad \Delta\equiv n^\mu\partial_\mu.
\eeq
Furthermore, one introduces a set of spin coefficients (Ricci rotations symbols),
\begin{eqnarray}
\alpha&=&-\textstyle{\frac{1}{2}}\left(n_{\mu;\nu}l^\mu\bar{m}^\nu-\bar{m}_{\mu;\nu}m^\mu\bar{m}^\nu\right),\\
\beta&=&\textstyle{\frac{1}{2}}\left(l_{\mu;\nu}n^\mu m^\nu-m_{\mu;\nu}\bar{m}^\mu m^\nu\right),\\
\gamma&=&-\textstyle{\frac{1}{2}}\left(n_{\mu;\nu}l^\mu n^\nu-\bar{m}_{\mu;\nu}m^\mu m^\nu\right),\\
\epsilon&=&\textstyle{\frac{1}{2}}\left(l_{\mu;\nu}n^\mu l^\nu-m_{\mu;\nu}\bar{m}^\mu l^\nu\right),\\
\kappa&=&l_{\mu;\nu}m^\mu l^\nu, \qquad ~~\lambda=-n_{\mu;\nu}\bar{m}^\mu \bar{m}^\nu,\\
\rho&=&l_{\mu;\nu}m^\mu \bar{m}^\nu, \qquad \mu=-n_{\mu;\nu}\bar{m}^\mu m^\nu,\\
\sigma&=&l_{\mu;\nu}m^\mu m^\nu, \qquad \nu=-n_{\mu;\nu}\bar{m}^\mu n^\nu,\\
\tau&=&l_{\mu;\nu}m^\mu n^\nu, \qquad ~\pi=-n_{\mu;\nu}\bar{m}^\mu l^\nu.
\end{eqnarray}
Despite a seemingly large number of variables we will find this notation
very useful and practical later on. However, first it will be useful to
give an explicit example.

\subsubsection*{Example of the null tetrad for Schwarzschild metric}
The metric is written in the form
\beq
\rd s^2=\left(1-\frac{2M}{r}\right)\rd t^2-\left(1-\frac{2M}{r}\right)^{-1}\rd r^2-r^2\rd\theta^2-r^2\sin^2\theta\rd\phi^2.
\eeq
The appropriate null tetrad is then given by
\begin{eqnarray}
l^\mu&=&\left([1-2M/r]^{-1},1,0,0\right),\\
n^\mu&=&\left(\textstyle{\frac{1}{2}},\textstyle{\frac{1}{2}}[1-2M/r],0,0\right),\\
m^\mu&=&\frac{1}{\sqrt{2}\,r}\left(0,0,1,\Im\sin^{-1}\theta\right).
\end{eqnarray}
An arbitrary type-D spacetime (e.g.\ the Schwarszchild metric) allows to
set $\kappa=\sigma=\nu=\lambda=0$. In particular, for the Schwarzschild
metric the explicit form of non-vanishing spin coefficients is:
\beq
\rho=-\frac{1}{r},\quad \mu=-\frac{1}{2r}\frac{1}{1-2M/r},\quad
\alpha=-\beta=-\sqrt{2}\,r\cot\frac{\theta}{2},\quad \gamma=\frac{M}{2r^2}.
\eeq

\subsubsection*{Maxwell's equations}
Maxwell's equations adopt the form
\begin{eqnarray}
(D-2\rho+2\epsilon)\Phi_1-(\bar{\delta}+\pi-2\alpha)\Phi_0&=&2\pi J_l,\\
(\delta-2\tau)\Phi_1-(\Delta+\mu-2\gamma)\Phi_0&=&2\pi J_m,\\
(D-\rho+2\epsilon)\Phi_2-(\bar{\delta}+2\pi)\Phi_1&=&2\pi J_{\bar{m}},\\
(\delta-\tau+2\beta)\Phi_2-(\Delta+2\mu)\Phi_1&=&2\pi J_n
\end{eqnarray}
with
\begin{eqnarray}
J_l&=&l_\mu(j^\mu+\Im\mathcal{M}^\mu),\\
J_m&=&m_\mu(j^\mu+\Im\mathcal{M}^\mu),\\
J_{\bar{m}}&=&\bar{m}_\mu(j^\mu+\Im\mathcal{M}^\mu),\\
J_n&=&n_\mu(j^\mu+\Im\mathcal{M}^\mu).
\end{eqnarray}
These are four equations for three complex variables.

\subsubsection*{Teukolsky's equations} 
Teukolsky (1973) derived the following form of Maxwell equations:
\begin{eqnarray}
\Big[(D&\!\!-\!\!&\epsilon+\bar{\epsilon}-2\rho-\bar{\rho})(\Delta+\mu-2\gamma)\nonumber\\ 
&&-(\delta-\beta-\bar{\alpha}-2\tau+\bar{\pi})(\bar{\delta}+\pi-2\alpha\Big]\Phi_0=2\pi J_0,\\
\Big[(D&\!\!+\!\!&\epsilon+\bar{\epsilon}-\rho-\bar{\rho})(\Delta+2\mu)\nonumber\\
&&-(\delta+\beta-\bar{\alpha}-\tau+\bar{\pi})(\bar{\delta}+2\pi\Big]\Phi_1=2\pi J_1,\\
\Big[(\Delta&\!\!+\!\!&\gamma-\bar{\gamma}+2\mu+\bar{\mu})(D-\rho+2\epsilon)\nonumber\\
&&-(\bar{\delta}+\alpha+\bar{\beta}-\bar{\tau}+2\pi)(\delta-\tau+2\beta\Big]\Phi_2=2\pi J_2
\end{eqnarray}
with
\begin{eqnarray}
J_0&=&(\delta-\beta-\bar{\alpha}-2\tau+\bar{\pi})J_l-(D-\epsilon+\bar{\epsilon}-2\rho-\bar{\rho})J_m,\\
J_1&=&(\delta+\beta-\bar{\alpha}-\tau+\bar{\pi})J_{\bar{m}}-(D+\epsilon+\bar{\epsilon}-\rho-\bar{\rho})J_n,\\
J_2&=&(\Delta+\gamma-\bar{\gamma}+2\mu+\bar{\mu})J_{\bar{m}}-(\bar{\delta}+\alpha+\bar{\beta}+2\pi-\bar{\tau})J_n.
\end{eqnarray}
Clearly this is an extremely useful form: noticed that the above-given
differential equations are entirely decoupled.

\subsubsection*{Example -- Maxwell's equations in Schwarzschild metric}
\begin{eqnarray}
\left[\frac{\partial}{\partial r}+\frac{2}{r}\right]\Phi_1+\frac{1}{\sqrt{2}r} {^\star}\bar{\partial}\Phi_0&=&2\pi J_l,\\
-\frac{1}{\sqrt{2}r} {^\star}{\partial}\Phi_1+\frac{1}{2}\left[\left(1-\frac{2M}{r}\right)\frac{\partial}{\partial r}+\frac{1}{r}\right]\Phi_0&=&2\pi J_m,\\
\left[\frac{\partial}{\partial r}+\frac{1}{r}\right]\Phi_2+\frac{1}{\sqrt{2}r} {^\star}\bar{\partial}\Phi_1&=&2\pi J_{\bar{m}},\\
-\frac{1}{\sqrt{2}r} {^\star}{\partial}\Phi_2+\frac{1}{2}\left(1-\frac{2M}{r}\right)\left[\frac{\partial}{\partial r}+\frac{2}{r}\right]\Phi_1&=&2\pi J_n,
\end{eqnarray}
where the ``edth'' operator acts on a spin weight $s$ quantity $\eta$ is the following manner:
\beq
^\star\partial\eta=-\left\{\sin^s\theta\left[\frac{\partial}{\partial\theta}+\frac{\Im}{\sin\theta}\frac{\partial}{\partial\phi}\right]\sin^{-s}\theta\right\}\eta.
\eeq
Spin weight is defined by by the transformation property $\eta\rightarrow e^{\Im s\psi}\eta$ under the transformation $m\rightarrow e^{\Im\psi}m$.
$\Phi_0$, $\Phi_1$, $\Phi_2$ have spin weights $s=1$, $0$, $-1$, respectively.

\subsubsection*{Spin harmonics}
Spin harmonics form a complete set of orthonormal functions
\beq
_sY_{lm}(\theta,\phi)=\left\{
\begin{array}{ll}
\sqrt{\frac{(l-s)!}{(l+s)!}}\,^{\star}\partial^sY_{lm}(\theta,\phi)&\mbox{~for~}0\leq s\leq l,\\
(-1)^s\sqrt{\frac{(l+s)!}{(l-s)!}}\,^{\star}\partial^{-s}Y_{lm}(\theta,\phi)&\mbox{~for~}-l\leq s\leq 0
\end{array}\right.
\eeq
with the orthogonality relation
\beq
{\int_0^{2\pi}\int_0^{\pi}}{~_s}Y_{lm}(\theta,\phi)\; _sY_{l^\prime m^\prime}(\theta,\phi)\,\sin\theta\rd\theta\rd\phi=\delta_{ll^\prime}\delta_{mm^\prime}.
\eeq
A general stationary vacuum electromagnetic test field can be expanded
in terms of spin-$s$ spherical harmonics.

\subsection{Test fields in Schwarzschild spacetime}
Bi\v{c}\'ak \& Dvo\v{r}\'ak (1980) use the following expansion:
\begin{eqnarray}
\Phi_0&=&{\sum_{l=1}^\infty}{\sum_{m=-l}^l} {^0}R_{lm}(r)\,_1Y_{lm}(\theta,\phi),\\
\Phi_1&=&{\sum_{l=0}^\infty}{\sum_{m=-l}^l} {^1}R_{lm}(r)\,_0Y_{lm}(\theta,\phi),\\
\Phi_2&=&{\sum_{l=1}^\infty}{\sum_{m=-l}^l} {^2}R_{lm}(r)\,_{-1}Y_{lm}(\theta,\phi).
\end{eqnarray}

\begin{figure}[tbh]
\begin{center}
\includegraphics[width=0.45\textwidth]{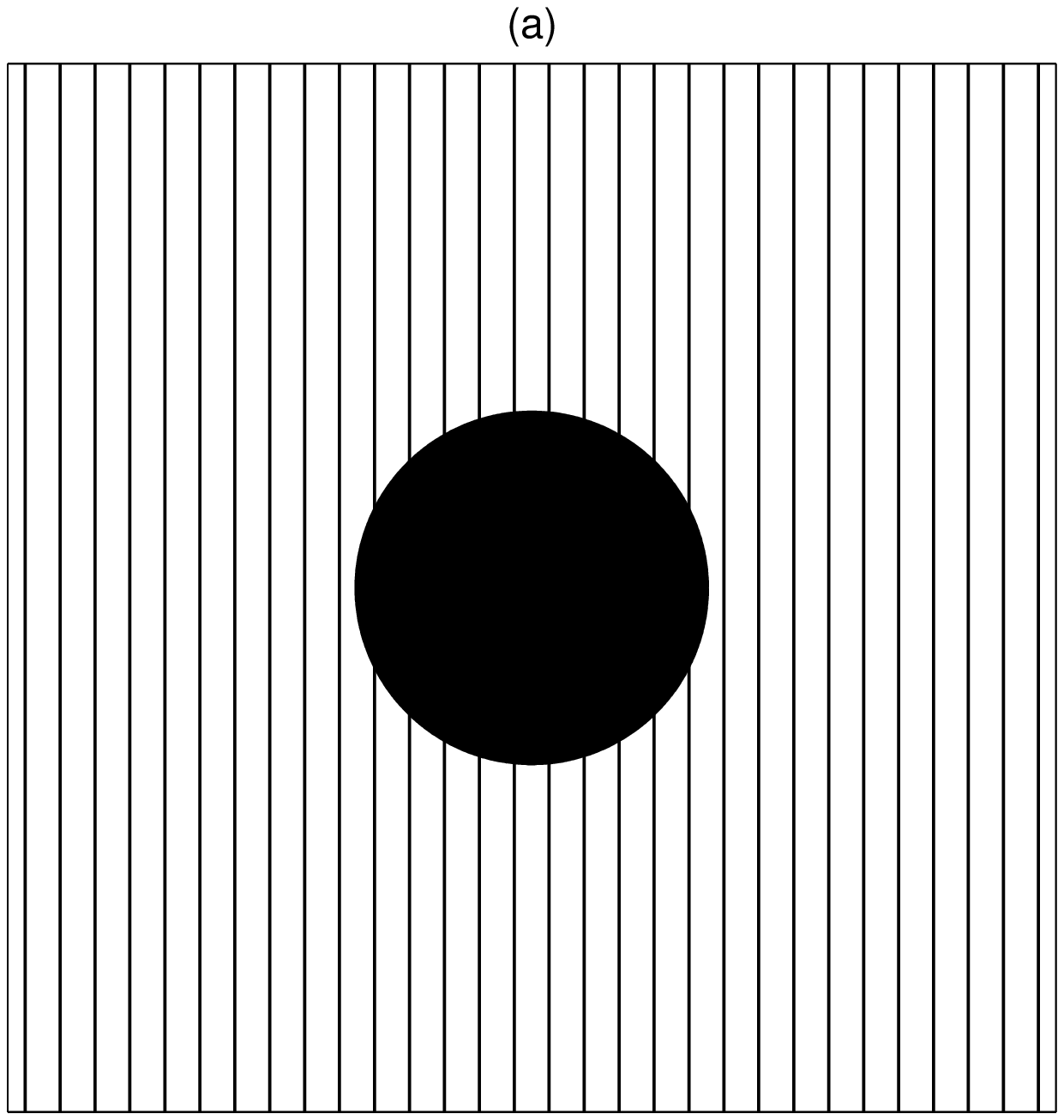}\hfill\includegraphics[width=0.45\textwidth]{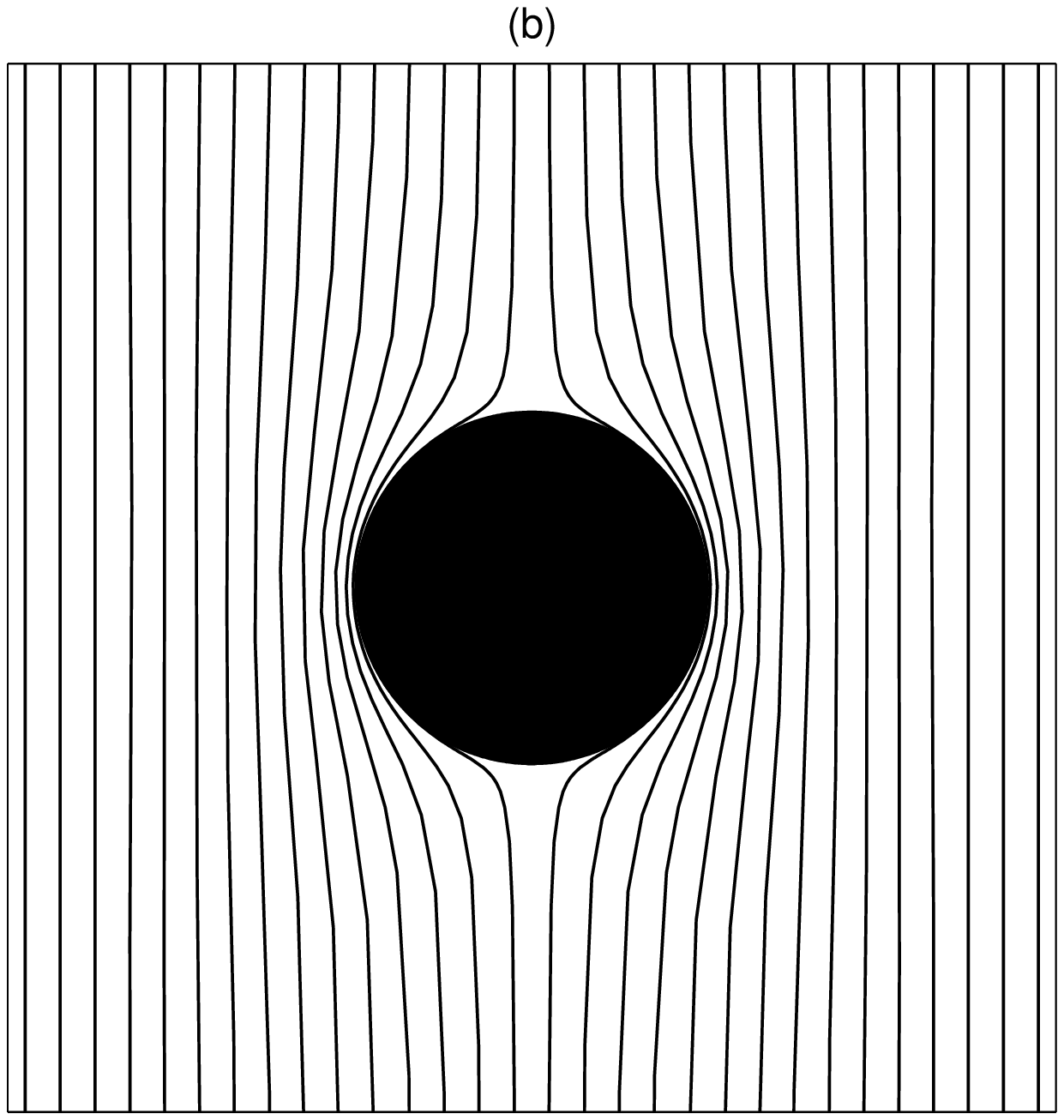}
\caption{An axisymmetric case: (a)~$a=0$ (static black hole), and (b)~$a=M$ (maximally rotating black hole).}
\label{fig1}
\end{center}
\end{figure}

Then the equations for radial functions take a form
\begin{eqnarray}
\hspace*{-2em}r(r-2M)\,^0R_{lm}^{\prime\prime}+4(r-M)\,^0R_{lm}^{\prime}-(l-1)(l+2)\,^0R_{lm}&\!=\!&-4\pi\,^0J_{lm},\\[1em]
\hspace*{-2em}r(r-2M)\,^1R_{lm}^{\prime\prime}+2(2r-3M)\,^1R_{lm}^{\prime}-(l-1)(l+2)\,^1R_{lm}&\!=\!&-4\pi\,^1J_{lm},\\[1em]
\hspace*{-2em}r(r-2M)\,^2R_{lm}^{\prime\prime}+4(r-2M)\,^2R_{lm}^{\prime}~~~~~~~~~~~~~~~~~~~~~~~~~~~~~ &&\\
-[(l-1)(l+2)+4M/r]\,^2R_{lm}&\!=\!&-4\pi\,^2J_{lm},
\end{eqnarray}
where
\begin{eqnarray}
^0J_{lm}(r)&=&\int J_0(r,\theta,\phi)\,_1\bar{Y}_{lm}(\theta,\phi)\,r^2\rd\Omega,\\
^1J_{lm}(r)&=&\int J_1(r,\theta,\phi)\,_0\bar{Y}_{lm}(\theta,\phi)\,r^2\rd\Omega,\\
^2J_{lm}(r)&=&\int J_2(r,\theta,\phi)\,_{-1}\bar{Y}_{lm}(\theta,\phi)\,r^2\rd\Omega.
\end{eqnarray}
A vacuum field solution is given by a Fuchsian-type equation (Bi\v{c}\'ak \& Dvo\v{r}\'ak 1980)
\beq
x(x-1)\frac{\rd^2\,^1R_{lm}}{\rd x^2}+(4x-3)\frac{\rd\,^1R_{lm}}{\rd x}-(l-1)(l+2)\,^1R_{lm}=0,
\eeq
with $x\equiv r/(2M)$.

\begin{figure}[tbh]
\begin{center}
\includegraphics[width=0.45\textwidth]{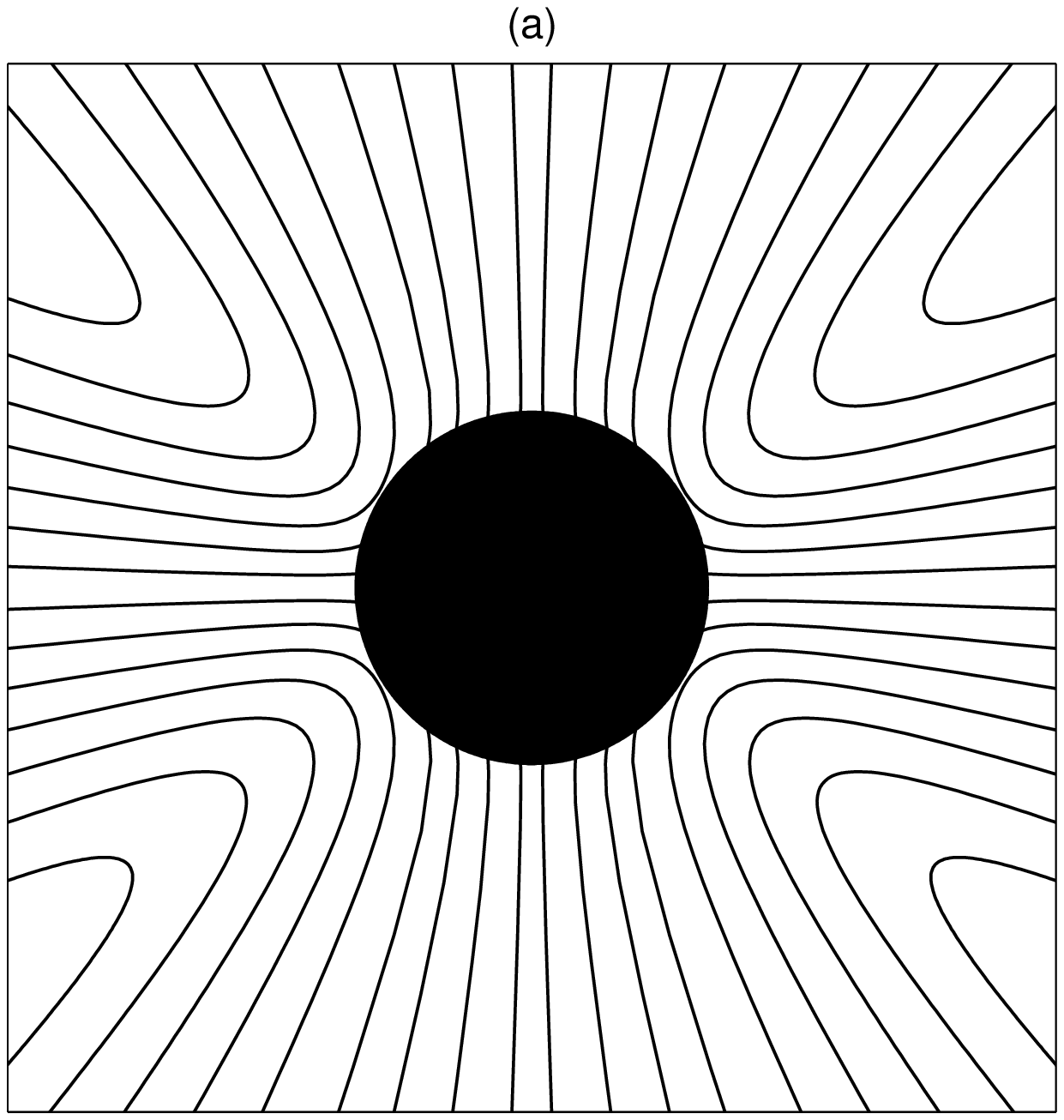}\hfill\includegraphics[width=0.45\textwidth]{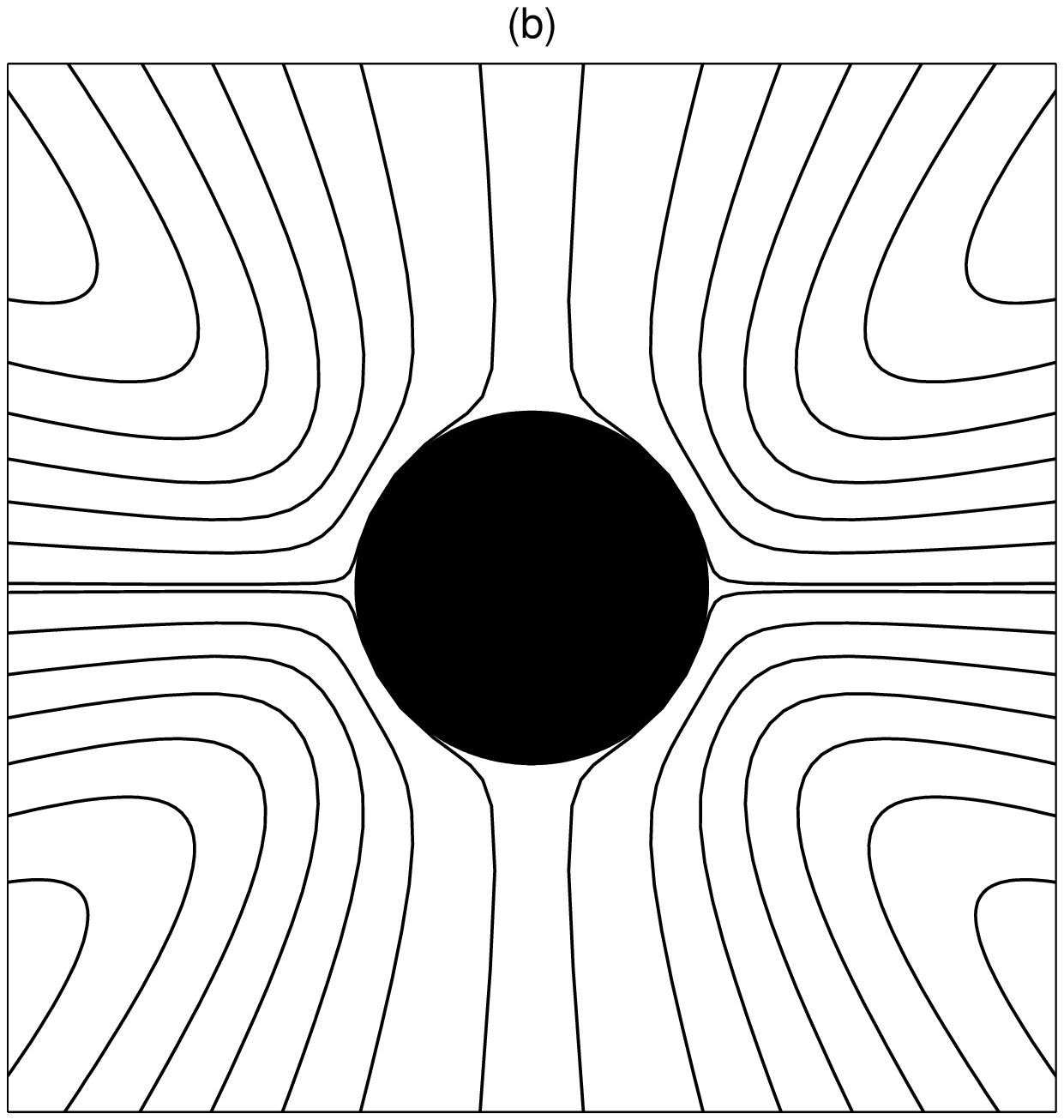}
\caption{The case of (a)~uniform aligned magnetic field near a fast rotating black hole ($a=0.95M$); (b)~near the maximally rotating hole ($a=M$).}
\label{fig2}
\end{center}
\end{figure}

Two independent solutions can be found:
\begin{eqnarray}
\left.\begin{array}{ll}
^1R^{(I)}_l&=F(1-l,l+2,3;x),\\ 
^1R^{(II)}_l&=\left(-x\right)^{-l-2}F(l,l+2,2l+2;x^{-1})
\end{array}\right\}\quad\mbox{for~}l\neq0,
\end{eqnarray}
\begin{eqnarray}
\left.\begin{array}{ll}
^1R^{(I)}_0&=\frac{1}{x^2}\ln(x-1)+\frac{1}{x}\\ 
^1R^{(II)}_0&=\frac{1}{x^2}
\end{array}\right\}\quad\mbox{for~}l=0.
\end{eqnarray}

A general solution reads
$^1R_{lm}=a_{lm}\,^1R^{(I)}_l+b_{lm}\,^1R^{(II)}_l$, $a_{lm},\,b_{lm}=\mbox{const.}$
Inserting the solution for $^1R_{lm}$ in Maxwell equations Bi\v{c}\'ak \& Dvo\v{r}\'ak 
(1980) find
\begin{eqnarray}
^0R_{lm}&=&a_{lm}\,^0R^{(I)}_l+b_{lm}\,^0R^{(II)}_l\;=\;\sqrt{\frac{2}{l(l+1)}}\frac{1}{r}\frac{\rd}{\rd r}\left(r^2\,^1R_{lm}\right),\\
^2R_{lm}&=&a_{lm}\,^2R^{(I)}_l+b_{lm}\,^2R^{(II)}_l,
\end{eqnarray}
where
\begin{eqnarray}
&&^0R^{(I)}_l=\frac{2\sqrt{2}}{\sqrt{l(l+1)}}\,F(1-l,l+2,2;x),\\
&&^0R^{(II)}_l=-\sqrt{\frac{2l}{l+1}}\,(-x)^{-l-2}F(l+1,l+2,2l+2;x^{-1}),\\
&&^2R^{(I)}_l=-\sqrt{\frac{2}{l(l+1)}}\,x^{-1}F(-l,l+1,2;x),\\
&&^2R^{(II)}_l=-\sqrt{\frac{l}{2(l+1)}}\,(-x)^{-l-2}F(l+1,l,2l+2;x^{-1}).
\end{eqnarray}
We can select a physically appropriate solution by assuming a source between $r_1$ and $r_2$ ($r_+\leq r_1\leq r_2\leq\infty$). By seeking a well-behaved solution on horizon that vanishes 
at infinity, we find

\begin{figure}[tbh]
\begin{center}
\includegraphics[width=0.45\textwidth]{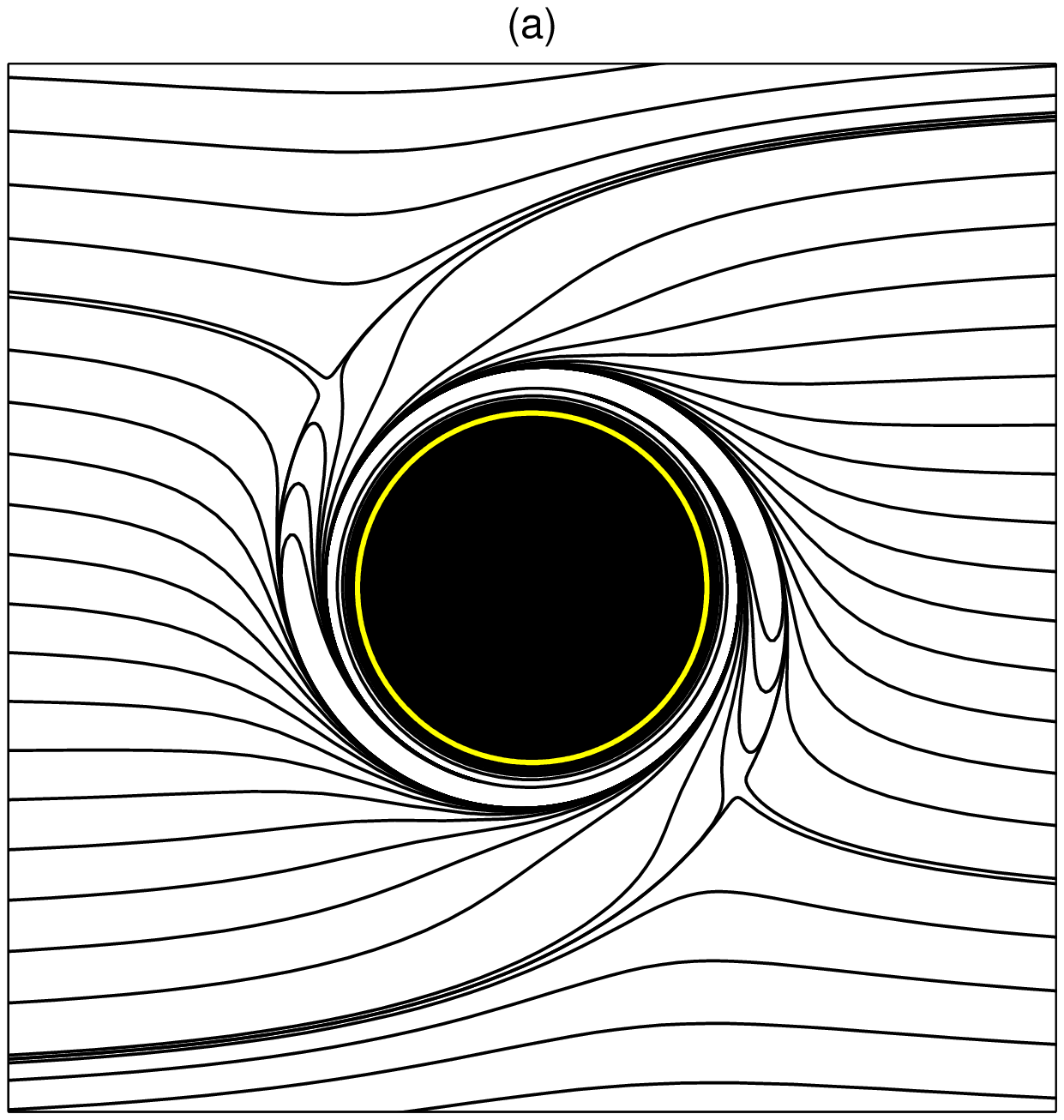}\hfill\includegraphics[width=0.45\textwidth]{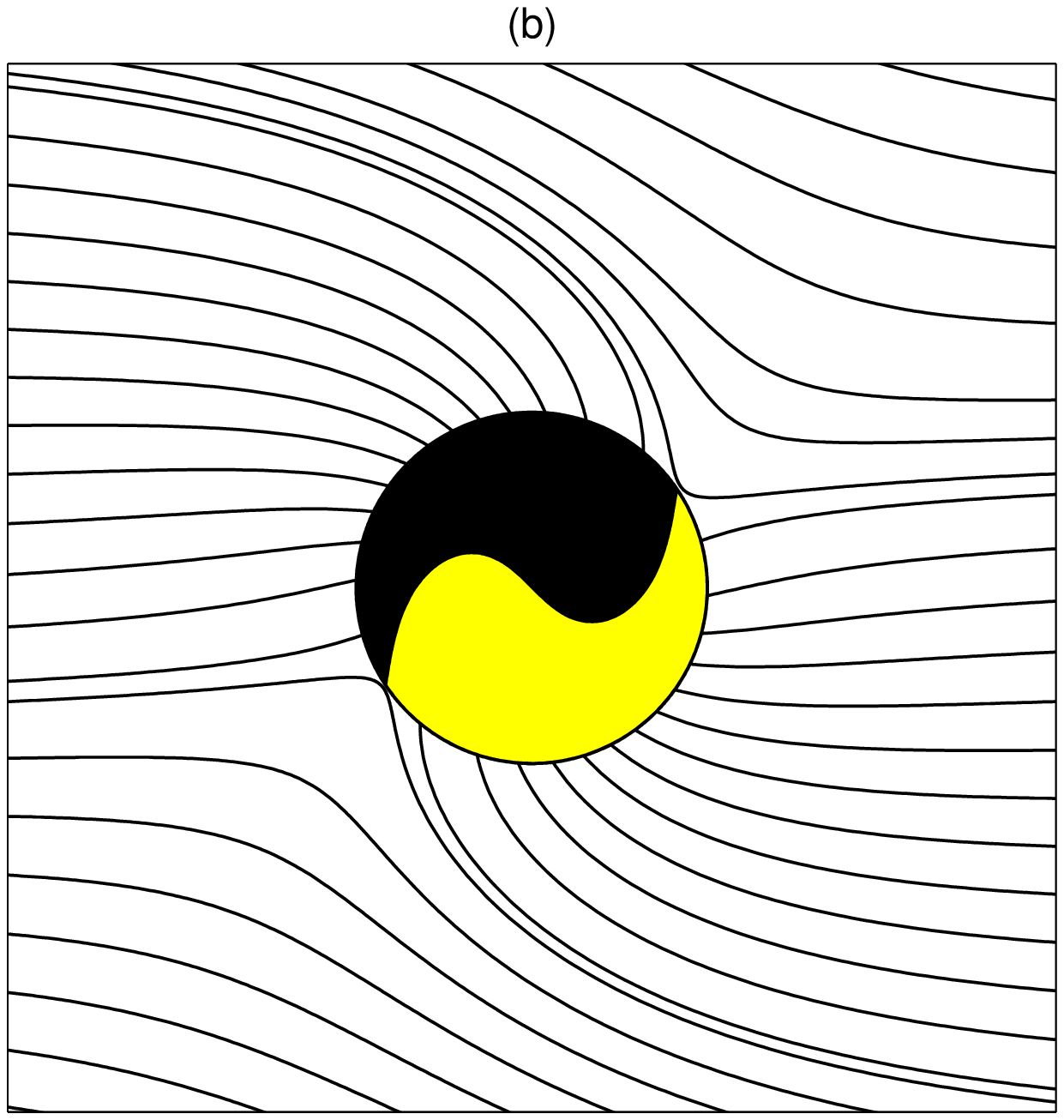}
\caption{Equatorial plane is shown as viewed from top, i.e.\ along rotation axis, (a)~in the frame of zero angular momentum observers orbiting at constant radius; (b)~in the frame of freely falling observers. In the panel (b), two regions of ingoing/outgoing lines are distinguished by different levels of shading of the horizon. The hole rotates counter-clockwise ($a=M$). Based on Karas (1989), Dov\v{c}iak et al. (2000).}
\label{fig3}
\end{center}
\end{figure}

\begin{eqnarray}
\left.\begin{array}{ll}
\Phi_0&=\sum_{l,m}a_{l,m}\,^0R^{(I)}_l\,_1Y_{lm}\\ 
\Phi_1&=\sum_{l,m}a_{l,m}\,^1R^{(I)}_l\,_0Y_{lm}+\frac{E_{\rm a}}{r^2}\,_0Y_{00}\\
\Phi_2&=\sum_{l,m}a_{l,m}\,^2R^{(I)}_l\,_{-1}Y_{lm}
\end{array}\right\}\quad\mbox{for~}2M\leq r<r_1,
\end{eqnarray}

\begin{eqnarray}
\left.\begin{array}{ll}
\Phi_0&=\sum_{l,m}b_{l,m}\,^0R^{(II)}_l\,_1Y_{lm}\\ 
\Phi_1&=\sum_{l,m}b_{l,m}\,^1R^{(II)}_l\,_0Y_{lm}+\frac{E_{\rm b}}{r^2}\,_0Y_{00}\\
\Phi_2&=\sum_{l,m}b_{l,m}\,^2R^{(II)}_l\,_{-1}Y_{lm}
\end{array}\right\}\quad\mbox{for~}r>r_2.\mbox{~~~~~~~~}
\end{eqnarray}

\subsubsection*{Two examples}
First, a spherically symmetric electric field. A unique solution that is
well-behaving both at $r=r_+$ and at $r\rightarrow\infty$:
$^1R^{(II)}_0$.  This term describes a weakly charged
Reissner-Nordstr\"om black hole.

Second, an asymptotically uniform magnetic field: 
\begin{eqnarray}
F_{\mu\nu}&\rightarrow&B_0\mbox{\boldmath$e_z$}+B_1\mbox{\boldmath$e_x$},\\
&\mbox{i.e.}&F_{r\theta}\rightarrow-B_1\,r\sin\phi,\\
&&F_{r\phi}\rightarrow B_0\,r\,\sin^2\theta-B_1\,r\,\sin\theta\,\cos\theta\,\cos\phi,\\
&&F_{\theta\phi}\rightarrow B_0\,r^2\,\sin\theta\cos\theta+B_1\,r^2\,\sin^2\theta\,\cos\phi.
\end{eqnarray}

\subsection{Magnetic and electric lines of force near a rotating black hole}
Lorentz force acts on electric/magnetic monopoles residing at rest with
respect to  a locally non-rotating frame,
\beq
\frac{\rd u^\mu}{\rd\tau}\propto \,^{\star}\!F^\mu_\nu\,u^\nu,\qquad \frac{\rd u^\mu}{\rd\tau}\propto F^\mu_\nu\,u^\nu.
\eeq
Magnetic lines are defined (Christodoulou \& Rufini 1973):
\beq
\frac{\rd r}{\rd\theta}=-\frac{F_{\theta\phi}}{F_{r\phi}},\qquad \frac{\rd r}{\rd\phi}=\frac{F_{\theta\phi}}{F_{r\theta}}.
\eeq
In an axially symmetric case the magnetic flux is:
\beq
\Phi_{\rm m}=\pi B_0\left[r^2-2Mr+a^2+\frac{2Mr}{r^2+a^2\,\cos^2\!\theta}\left(r^2-a^2\right)\right]\sin^2\!\theta=\mbox{const}.
\eeq
Notice: $\Phi_{\rm m}=0$ for $r=r_+$ and $a=M$. The flux is expelled out of the horizon 
(Meissner effect; Bi\v{c}\'ak \& Ledvinka 2000; Penna 2014).

The electric fluxes and field lines can be introduced in a similar
manner, one only needs to interchange the electromagnetic field tensor
by its dual, the magnetic charge by the electric charge, and vice versa
wherever they appear in the above-given formulae. It should be evident
that the induced electric field vanishes in the non-rotating case. Based
on the classical analogy with a rotating sphere, one would perhaps
expect a quadrupole-type component, but here the leading term of the
electric field arises due to gravomagnetic interaction which is a purely
general-relativistic effect, and this electric field falls off radially
as $r^{-2}$.

Magnetic field lines reside in surfaces of constant magnetic flux, and
this way the lines of force are defined in an invariant way (see
Fig.~\ref{fig1}). Electric field is induced by the gravito-magnetic
influence of the  black hole. The resulting field lines are shown in
Fig.~\ref{fig2}. An asymptotic form of the electric field-lines reads
\begin{eqnarray}
\frac{\rd r}{\rd\lambda}&=&\frac{B_0aM}{r^2}\left(3\cos^2\theta-1\right)\;+\;\frac{3B_{\perp}aM}{r^2}\,\sin\theta\,\cos\theta\,\cos\phi+\mathcal{O}\left(r^{-3}\right),\\
\frac{\rd\theta}{\rd\lambda}&=&\mathcal{O}\left(B_{\perp}r^{-3}\right),\qquad\frac{\rd\phi}{\rd\lambda}\;=\;\mathcal{O}\left(B_{\perp}r^{-3}\right).
\end{eqnarray}
As mentioned above, an aligned magnetic field produces an asymptotically
radial electric field, rather than a quadrupole field, expected under
these circumstances in the classical electrodynamics. This difference is
due to rotation of the black hole.

Fig.~\ref{fig3} shows the structure of a uniform magnetic field
perpendicular to the black hole rotation axis (Bi\v{c}\'ak \& Karas
1989; Karas et al.\ 2009, 2012, 2013, 2014). We notice the enormous effect of frame-dragging which acts on
field lines and distorts them in the sense of black hole rotation.
Nevertheless, some field lines still enter the horizon and bring the
magnetic flux into the black hole (naturally, the same magnetic flux has
to emerge out of the horizon, so that the total flux through the black
hole vanishes and its magnetic charge is equal zero). 

\begin{figure}[tbh]
\begin{center}
\includegraphics[width=0.9\textwidth]{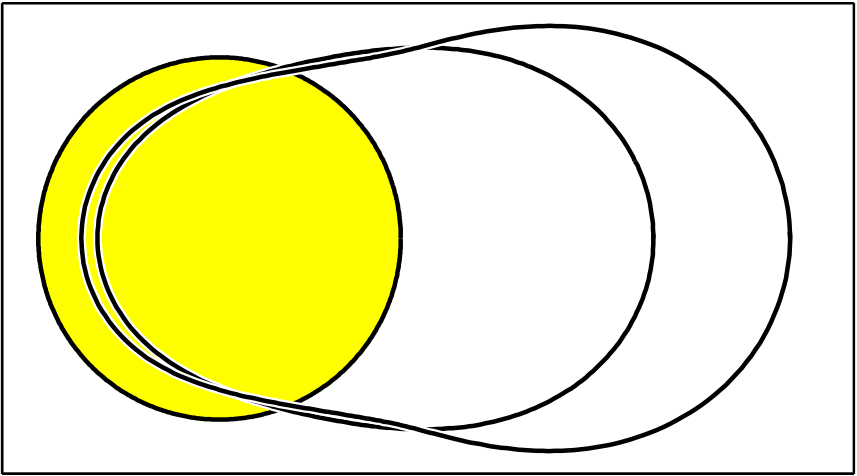}
\caption{Cross-sectional area for the capture of magnetic flux by a rotating black hole. 
The three curves correspond to different values of the black-hole
angular momentum: $a=0$ (cross-section is the circle; its projection
coincides with the black-hole horizon, indicated by yellow colour),
$a=0.95\,M$, and $a=M$. The enclosed area contains the field lines of 
the asymptotically perpendicular magnetic field whiech eventually enter
into the black hole horizon. From the graph we notice that this area
grows with the black hole spin and its shape is distorted by the
gravitomagnetic interaction.}
\label{fig4}
\end{center}
\end{figure}

We notice that magnetic null points emerge near the black hole,
suggesting that magnetic reconnection can be initiated by the purely
gravitomagnetic effect of the rotating black hole. Indeed, this new
reconnection mechanism has been only recently proposed (Karas \&
Kop\'a\v{c}ek 2009) in the context of particle acceleration processes
near magnetized black holes. The capture of magnetic field lines is
further illustrated in Fig.~\ref{fig4} where we plot the black hole
effective cross  sectional area.

\subsubsection*{Surface charge on the horizon}
Surface charge is formally defined by the radial component of electric field in non-singular coordinates (Thorne et al.\ 1986),
\begin{eqnarray}
\sigma_{\mathrm{H}}&=&\frac{B_0a}{4\pi\Sigma_+}\left[r_+\sin^2\theta-\frac{M}{\Sigma_+}\left(r_+^2-a^2\cos^2\theta\right)\left(1+\cos^2\theta\right)\right]\\
&&+\frac{B_{\perp}a}{4\pi\Sigma_+}\,\sin\theta\,\cos\theta\left[\frac{Mr_+}{\Sigma_+}+1\right]\left[a\sin\psi-r_+\cos\psi\right],
\end{eqnarray}
with
\beq
\psi=\phi+\frac{a}{r_+-r_-}\,\ln\frac{r-r_+}{r-r_-}\,\propto\ln(r-r_+).
\eeq
For $a\ll M$,
\beq
\sigma_{\mathrm{H}}=\frac{a}{16\pi M}\left[B_0\left(1-3\cos^2\theta\right)+3B_{\perp}\sin\theta\,\cos\theta\,\cos\psi\right].
\eeq
It should be obvious that $\sigma_{\mathrm{H}}$ does not represent any
kind of a real charge distribution. Instead, it is introduced only by
pure analogy with junction conditions for Maxwell's equations in
classical electrodynamics. The
classical problem was treated in original works by Faraday, Lamb,
Thomson and Hertz, and more recently in Bullard (1949) and Elsasser (1950). 
It is quite enlightening to pursue this similarity
to greater depth (see e.g.\ Karas \& Bud\'{\i}nov\'a 2000, and references cited
therein) despite the fact that this is purely a formal analogy,
as pointed out by Punsly (2008).


\section{On the way from test fields to exact solutions of Einstein-Maxwell equations}
So far we discussed test-field solutions of Einstein equations which reside in a
prescribed (curved) spacetime. In the rest of this lecture we will briefly outline a way
to construct {\em exact\/} solutions of mutually couple (vacuum) Einstein-Maxwell
equations. Because this task is very complicated, astrophysically realistic results
can be only obtained by numerical approaches. However, important insight can be
gained by simplified analytic solutions. We will thus explore the latter approach. 

\subsubsection*{The spacetime metric}
Let us first assume a static spacetime metric in the form
\beq
\rd s^2=f^{-1}\left[e^{2\gamma}\left(\rd z^2+\rd\rho^2\right)+\rho^2\rd\phi^2\right]-f\left(\rd t-\omega\rd\phi\right)^2,
\eeq
with $f$, $\omega$, and $\gamma$ being functions of $z$ and $\rho$ only.
We consider coupled Einstein-Maxwell equations under the following constraints:
(i)~electrovacuum case containing a black hole, (ii)~axial symmetry and stationarity,
(iii) {\it not} necessarilly asymptotically flat (see Kramer et al. 1980; Alekseev \&
Garcia 1996; Ernst \& Wild 1976; Karas \& Vokrouhlick\'y 1990, and references 
cited therein). 

As explained in various textbooks and, namely, in the above-mentioned works, 
one can proceed in the 
following way to find the three unknown metric functions:
\begin{itemize}
\item Standard approach: $g_{\mu\nu}\rightarrow\Gamma^\mu_{\nu\lambda}\rightarrow R^\alpha_{\beta\gamma\delta}\rightarrow G_{\mu\nu}$
\item Exterior calculus: $e^\mu_{(\lambda)}\rightarrow\omega_{\mu\nu}\Omega_{\mu\nu}\rightarrow R^{\hat{\alpha}}_{\hat{\beta}\hat{\gamma}\hat{\delta}}\rightarrow G_{\hat{\mu}\hat{\nu}}$
\item Variation principle: $\mathcal{L}=-\frac{1}{2}\rho f^{-2}\nab f\!\cdot\!\nab f+\frac{1}{2}\rho^{-1}f^2\nab\omega\!\cdot\!\nab\omega$
\end{itemize}
We denoted nabla operator, $\nab\!\cdot\!\left(\rho^{-1}\vec{e_\phi\times}\nab\varphi\right)=0$ $\forall\varphi\equiv\varphi(\rho,z)$.
Now, the vacuum field equations (without electromagnetic field) can be written in the form:
\beq
f\nab^2f=\nab f\cdot\nab f-\rho^{-2}f^4\nab\omega\cdot\nab\omega,
\nab\cdot\left(\rho^{-2}f^2\nab\omega\right)=0.
\eeq
Let us define functions $\varphi(\rho,z)$, $\omega(\rho,z)$ by
the prescription
\begin{eqnarray*}
\rho^{-1}f^2\nab\omega&=&\vec{e_\phi\times}\nab\varphi,\\
f^{-2}\nab\varphi&=&-\rho^{-1}\vec{e_\phi\times}\nab\omega
\end{eqnarray*}
By applying $\nab\cdot$ operator on the both sides of the last equation,
the relation for $\varphi$ comes out,
$\nab\cdot\left(f^{-2}\nab\varphi\right)=0$.
Let us further define $\mathcal{E}\equiv f+\Im\varphi$. Then, both field equations can be written in the form
\beq
(\Re\mathcal{E})\nab^2\mathcal{E}=\nab\mathcal{E}\cdot\nab\mathcal{E}.
\eeq
Now we can proceed to adding the electromagnetic field:
\beq
\mathcal{L}^\prime=\mathcal{L}+2\rho f^{-1}A_0\left(\nab A\right)^2-2\rho^{-1}f\left(\nab A_3-\omega\nab A_0\right)^2
\eeq

Functions $f$, $\omega$, $A_0$, and $A_3$ are constrained by the variational
principle. Define $\Phi\equiv\Phi(A_0,A_3)$, $\mathcal{E}\equiv f-|\Phi|^2+\Im\varphi$:
\beq
\begin{array}{ll}
&\left(\Re\mathcal{E}+|\Phi|^2\right)\nab^2\mathcal{E}=\left(\nab\mathcal{E}+2\bar{\Phi}\nab\Phi\right)\cdot\nab\mathcal{E},\\
&\left(\Re\mathcal{E}+|\Phi|^2\right)\nab^2\Phi=\left(\nab\mathcal{E}+2\bar{\Phi}\nab\Phi\right)\cdot\nab\Phi.
\end{array}
\eeq
Let us assume $\mathcal{E}\equiv\mathcal{E}(\Phi)$ to be an analytic function
which satisfies
\beq
\left(\Re\mathcal{E}+\Phi^2\right)\frac{\rd^2\mathcal{E}}{\rd\Phi^2}\nab\Phi\cdot\nab\Phi=0.
\eeq
Assume further a linear relation,
\beq
\mathcal{E}=1-2\frac{\Phi}{q},\qquad q\in\mathrm{C}
\eeq
and a new variable $\xi$,
\beq
\mathcal{E}\equiv\frac{\xi-1}{\xi+1},\qquad\Phi=\frac{q}{\xi+1},
\eeq
\beq
[\xi\bar{\xi}-(1-q\bar{q})]\nab^2\xi=2\bar{\xi}\nab\xi\cdot\nab\xi.
\eeq

\subsubsection*{Generating ``new'' solutions}
We introduce new variables by relations
\beq
\xi_0\rightarrow\xi=(1-q\bar{q})\xi_0,
\eeq
\beq
[\xi_0\bar{\xi}_0-1]\nab^2\xi_0=2\bar{\xi}_0\nab\xi_0\cdot\nab\xi_0.
\eeq
i.e.
\beq
(\Re\mathcal{E}_0)\nab^2\mathcal{E}_0=\nab\mathcal{E}_0\cdot\nab\mathcal{E}_0,\qquad\mathcal{E}_0\equiv\frac{\xi_0-1}{\xi_0+1}.
\eeq
where $\mathcal{E}_0$ has a meaning of an ``old'' vacuum solution.

\textit{Theorem}. Let $(\Phi,\mathcal{E},\gamma_{\alpha\beta})$ be a solution of Einstein-Maxwell electrovaccum eqs. with anisotropic Killing vector field. Then there is another solution $(\Phi^\prime,\mathcal{E}^\prime,\gamma^\prime_{\alpha\beta})$, related to the old one by transformation
\begin{eqnarray*}
\mathcal{E}^\prime&=&\alpha\bar{\alpha}\mathcal{E},\quad\Phi^\prime=\alpha\Phi,\quad\mbox{\ldots dual rotation,~}^{\star}F_{\mu\nu}\rightarrow\sqrt{{\alpha}/{\bar{\alpha}}}\,^{\star}F_{\mu\nu},\\
\mathcal{E}^\prime&=&\mathcal{E}+\Im b,\quad\Phi^\prime=\Phi,\quad\mbox{\ldots calibration, no change in~}F_{\mu\nu},\\
\mathcal{E}^\prime&=&\mathcal{E}-2\bar{\beta}\Phi-\beta\bar{\beta},\quad\Phi^\prime=\Phi+\beta,\quad\mbox{\ldots calibration \ldots},\\
\mathcal{E}^\prime&=&\mathcal{E}(1+\Im c\mathcal{E})^{-1},\quad\Phi^\prime=(1+\Im c\mathcal{E})^{-1},\\
\mathcal{E}^\prime&=&\mathcal{E}{\underbrace{(1-2\bar{\gamma}\Phi-\gamma\bar{\gamma}\mathcal{E})}_{\Lambda=1-B_0\Phi-{\frac{1}{4}}B_0^2\mathcal{E}}}^{-1},\quad\Phi^\prime=(\Phi+\gamma\mathcal{E})(1-2\bar{\gamma}\Phi-\gamma\bar{\gamma}\mathcal{E})^{-1}.
\end{eqnarray*}
\beq
\mathcal{E}\rightarrow\mathcal{E}^\prime=\Lambda^{-1}\mathcal{E},\qquad f\rightarrow f^\prime=|\Lambda|^{-2}f,\qquad\omega\rightarrow\omega^\prime=\mbox{?},
\eeq
\beq
\Phi\rightarrow\Phi^\prime=\Lambda^{-1}(\Phi-\textstyle{\frac{1}{2}}B_0\mathcal{E}),\ \nab\omega^\prime=|\Lambda|^2\nab\omega+\rho f^{-1}(\bar{\Lambda}\nab\Lambda-\Lambda\nab\bar{\Lambda}).
\eeq

\subsubsection*{Examples}

\noindent
\textit{Example 1. Minkowski spacetime $\rightarrow$ Melvin universe.}
\beq
\rd s^2=\left[\rd z^2+\rd\rho^2-\rd t^2\right]+\rho^2\rd\phi^2.
\eeq
\beq
f=-\rho^2,\quad\omega=0,\quad\Phi=0,\quad\mathcal{E}=-\rho^2,\quad\varphi(\omega)=0.
\eeq
\beq
f^\prime=-\Lambda^{-2}\rho^2,\quad\omega^\prime=0,\quad\Phi^\prime=\textstyle{\frac{1}{2}}\Lambda^{-1}B_0\rho^2,
\eeq
\beq
B_z=\Lambda^{-2}B_0,\quad B_\rho=B_\phi=0,
\eeq
\beq
\rd s^2=\Lambda^2\left[\rd z^2+\rd\rho^2-\rd t^2\right]+\Lambda^{-2}\rho^2\rd\phi^2.
\eeq
Gravity of the magnetic field in balance with the Maxwell pressure. Cylindrical symmetry along $z$-axis.


\noindent
\textit{Example 2. Schwarzschild BH $\rightarrow$ Schwarzschild-Melvin black hole.}
\beq
\rd s^2=\left[\left(1-\frac{2M}{r}\right)^{-1}\rd r^2 - \left(1-\frac{2M}{r}\right)\rd t^2+r^2\rd\theta^2\right]+r^2\sin^2\theta\rd\phi^2,
\eeq
\beq
f=-r^2\sin^2\theta,\quad\omega=0,\quad\rho=\sqrt{r^2-2Mr}\,\sin\theta,
\eeq
\beq
B_r=\Lambda^{-2}B_0\cos\theta,\quad B_\theta=-\Lambda^{-2}B_0(1-2M/r)\sin\theta,
\eeq
\beq
\rd s^2=\Lambda^2\Big[ \quad...\quad\Big]+\Lambda^{-2}r^2\sin^2\theta\rd\phi^2.
\eeq
There the following limits of the magnetized Schwarzschild-Melvin black hole:
(i) $B_0=0$ ... Schwarzschild solution,
(ii) $r/M\rightarrow\infty$ ... Melvin solution,
(iii) $|B_0M|\ll1$ ... Wald's test field in the region $2M\ll r\ll B^{-1}_0$.


\noindent
\textit{Example 3. Magnetized Kerr-Newman BH.}
\begin{eqnarray*}
g&=&|\Lambda|^2\Sigma\left(\Delta^{-1}\rd{r}^2+\rd{\theta}^2-\Delta{A^{-1}}\rd{t}^2\right)\\
&&+|\Lambda|^{-2}\Sigma^{-1}A\sin^2\theta\left(\rd{\phi}-\omega\rd{t}\right)^2,
\label{g}
\end{eqnarray*}
$\Sigma=r^2+a^2\cos^2\theta$, $\Delta=r^2-2Mr+a^2+e^2$, $A=(r^2+a^2)^2-{\Delta}a^2\sin^2\theta$ are functions from the Kerr-Newman metric.

$\Lambda=1+\beta\Phi-\frac{1}{4}\beta^2\mathcal{E}$ is given in terms of the Ernst complex potentials $\Phi(r,\theta)$ and $\mathcal{E}(r,\theta)$:
\begin{eqnarray*}
\Sigma\Phi
 &=& ear\sin^2\theta-{\Im}e\left(r^2+a^2\right)\cos\theta, \\
\Sigma\mathcal{E}
 &=& -A\sin^2\theta-e^2\left(a^2+r^2\cos^2\theta\right)
 \nonumber \\
 & & + 2{\Im}a\left[\Sigma\left(3-\cos^2\theta\right)+a^2\sin^4\theta-
 re^2\sin^2\theta\right]\cos\theta.
\end{eqnarray*}


\begin{figure}
\begin{center}
\includegraphics[width=0.9\textwidth]{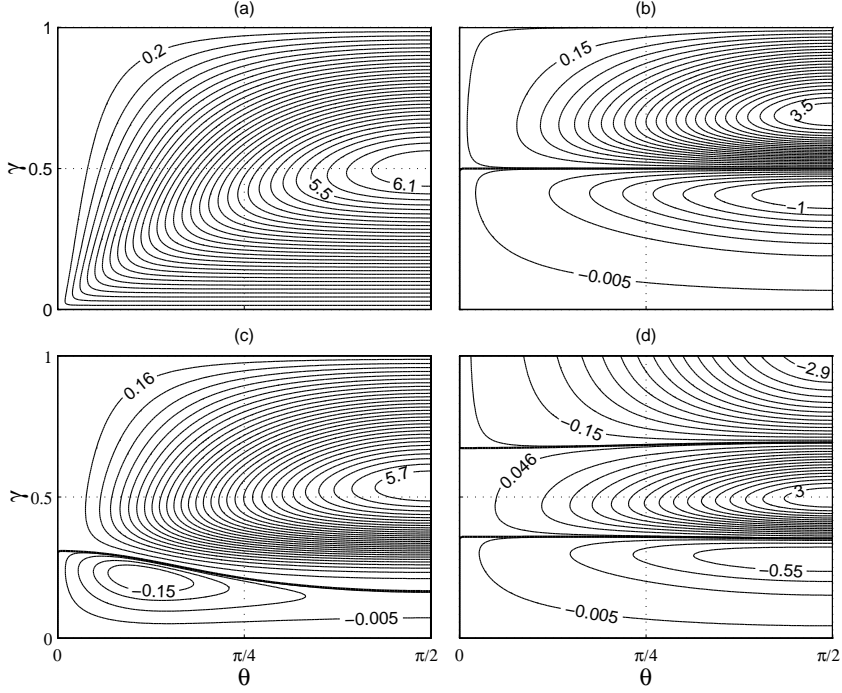}
\end{center}
\caption{Contours of magnetic flux across a cap on the horizon (latitude angle $\theta$ is measured from the rotation axis) of a magnetized black hole: 
(a)~$a=e=0$; (b)~$a=1$, $e=0$; (c)~$a=0.2$, $e=0$; (d)~$a=-e=1/\protect\sqrt{2}$ (electric charge and spin of the black hole). Here, $\gamma\equiv\left(1+\beta\right)^{-1}$, $\beta\equiv B_0M$. This figure from Karas \& Bud\'{\i}nov\'a (2000) illustrates strong-gravity effects on magnetic fields that do not occur in weak-magnetic (test) field approximation, namely, the expulsion of the magnetic flux as a function of the intensity of the imposed magnetic field.}
\label{fig5}
\end{figure}


The electromagnetic field can be written in terms of orthonormal LNRF components,
\begin{eqnarray*}
H_{(r)}+{\ri}E_{(r)} &=& A^{-1/2}\sin^{-1}\!\theta\,\Phi^{\prime}_{,\theta},\\
H_{(\theta)}+{\ri}E_{(\theta)} &=&-\left(\Delta/A\right)^{1/2}\sin^{-1}\!\theta\,\Phi^{\prime}_{,r},
\label{heth}
\end{eqnarray*}
where $\Phi^{\prime}(r,\theta)=\Lambda^{-1}\left(\Phi-\frac{1}{2}\beta\mathcal{E}\right)$.
\vspace*{2em}
The horizon is positioned at $r{\equiv}r_+=1+\sqrt(1-a^2-e^2)$, independent of $\beta$. As in the non-magnetized case, the horizon exists only for $a^2+e^2\leq1$.


There is an issue with this solution, namely, the 
range of angular coordinates \textit{versus} the problem of conical singularity: $0\leq\theta\leq\pi$, $0\leq\phi<2\pi|\Lambda_0|^2$, where 
\beq
|\Lambda_0|^2\equiv|\Lambda(\sin\theta=0)|^2= 1+\textstyle{\frac{3}{2}}\beta^2e^2+2\beta^3ae+\beta^4\left(\textstyle{\frac{1}{16}}e^4+a^2\right).
\eeq
\vspace*{2em}
The total electric charge $Q_{\rm{H}}$ and the magnetic flux $\Phi_{\rm{m}}(\theta)$ across a cap in axisymmetric position on the horizon (with the edge defined by $\theta={\rm{const}}$):
\begin{eqnarray*}
Q_{\rm{H}} &=& -|\Lambda_0|^2\,\Im{\rm{m}\,}\Phi^{\prime}\left(r_+,0\right),
\label{qh} \\
\Phi_{\rm{m}} &=& 2\pi|\Lambda_0|^2\,\Re{\rm{e}\,}\Phi^{\prime}
 \left(r_+,\bar{\theta}\right)\Bigr|\strut^{\theta}_{\bar{\theta}=0}.
\label{fh}
\end{eqnarray*}
The magnetic flux across the black hole hemisphere in the exact magnetized black hole
 solution is shown in fig. \ref{fig5}.


\ack
The author acknowledges continued support from the Czech Science 
Foundation grant No.\ 13-00070J.

\end{document}